\documentclass[twocolumn]{article}
\usepackage[a4paper, total={180mm, 243mm}]{geometry}
\usepackage[T1]{fontenc}
\usepackage[utf8]{inputenc}
\usepackage{lipsum}
\usepackage{graphicx}
\usepackage{siunitx}
\usepackage[version=4]{mhchem}
\usepackage{titling}
\usepackage{circledsteps}
\usepackage{subcaption}
\usepackage{mhchem}
\usepackage{amsmath, amssymb}
\usepackage{soul}

\DeclareCaptionLabelSeparator{custom}{\textbar}
\DeclareCaptionFormat{custom}{\textsf{\textbf{#1 #2} \small #3}}
\graphicspath{{./figures}} 

\makeatletter
\newcommand{\showfontsize}{\f@size{} pt}
\makeatother

\usepackage[
    backend=biber,
    style=nature,
    date=year,
    maxnames=99
]{biblatex}
\newcommand{\reffig}[2]{Fig.~\ref{#1}{\textbf{#2}}}

\newcommand\identity{1\kern-0.25em\text{l}}

\title{Quadrature squeezing in a nanophotonic microresonator}

\author{Alexander~E.~Ulanov$^{1,\dagger}$,
        Bastian Ruhnke$^{1}$,
        Thibault Wildi$^{1}$,
        Tobias Herr$^{1,2,*}$
}
\date{%
    \small $^1$Deutsches Elektronen-Synchrotron DESY, Notkestr. 85, 22607 Hamburg, Germany \\
    \small $^2$Physics Department, University of Hamburg UHH, Luruper Chaussee 149, 22761 Hamburg, Germany\\
    \small $^{\dagger}$alexander.ulanov@desy.de\\
    \small $^{*}$tobias.herr@desy.de
}

\begin{document}

\maketitle

\textbf{
Squeezed states of light are essential for emerging quantum technology in metrology and information processing. Chip-integrated photonics offers a route to scalable and efficient squeezed light generation, however, parasitic nonlinear processes and optical losses remain significant challenges. Here, we demonstrate single-mode quadrature squeezing in a photonic crystal microresonator via degenerate dual-pump spontaneous four-wave mixing. Implemented in a scalable, low-loss silicon-nitride photonic-chip platform, the microresonator features a tailored nano-corrugation that modifies its resonances to suppress parasitic nonlinear processes. In this way, we achieve an estimated 7.8~dB of on-chip squeezing in the bus waveguide, with potential for further improvement. These results open a promising pathway toward integrated squeezed light sources for quantum-enhanced interferometry, Gaussian boson sampling, coherent Ising machines, and universal quantum computing.
}

\subsection*{Introduction}

Squeezed light is an essential resource for emerging quantum technologies, including gravitational wave detection \cite{tse:2019}, quantum metrology \cite{giovannetti:2011}, imaging \cite{moreau:2019}, and quantum computing \cite{zhong:2020, madsen:2022}. 
It can be generated through nonlinear parametric processes and in one of its quadratures exhibits noise below that of classical coherent laser sources \cite{lvovsky:2015, andersen:2016, schnabel:2017}. 
Since the seminal demonstration of 0.3~dB noise reduction in hot sodium vapor in 1985 \cite{slusher:1985a}, the field has made significant progress, culminating in the current record of 15 dB achieved using a periodically poled KTP crystal cavity \cite{vahlbruch:2016}. 
In many systems, the squeezing process can be accompanied by competing nonlinear effects that can obstruct the observation of squeezing and hinder its practical utilization. Thus to enable effective generation of squeezed light, it is crucial to not only optimize the squeezing process itself but also to 
suppress unwanted nonlinear processes. Moreover, it is important to minimize optical losses as they degrade the level of squeezing.

Complementing squeezed light sources based on bulk optics, significant 
advances have been made in the development of photonic chip-integrated sources \cite{wang:2020} based on spontaneous parametric down-conversion in $\chi^{(2)}$ nonlinear waveguides, such as periodically poled thin-film lithium niobate \cite{kashiwazaki:2021, nehra:2022, chen:2022a, peace:2022, park:2024, arge:2024}, and four-wave mixing in $\chi^{(3)}$ Kerr-nonlinear materials \cite{dutt:2015, okawachi:2015, Vaidya2020, cernansky:2020, zhao:2020, zhang:2021, jahanbozorgi:2023, shen:2024b}.

Kerr-nonlinear systems are attractive for scalable quantum technologies as they provide access to ultra-low loss photonic-integrated waveguides and high-quality factor $Q$ optical microresonators ($Q>10$~million) \cite{ji:2017, liu:2021, ye:2022, zhang:2024}, enabling efficient excitation of nonlinear effects and squeezing. Importantly, they can also be produced in existing wafer-scale processes and are compatible with hybrid integration of laser sources and detectors.

\begin{figure*}[!ht]
  \centering
  \includegraphics[width=\textwidth]{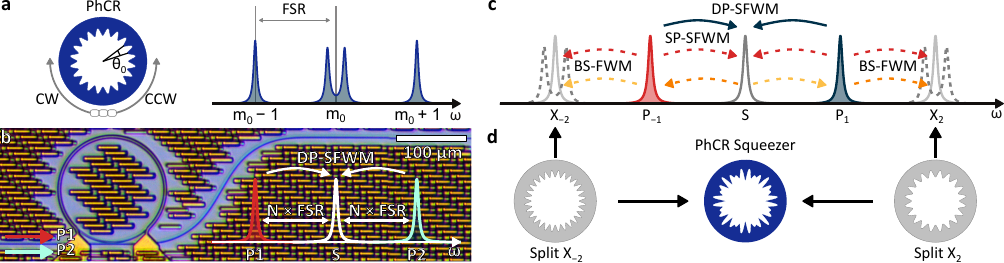}
  \caption{
    \textbf{Dual-pump spontaneous four-wave mixing (DP-SFWM) in a photonic crystal ring microresonator (PhCR)}. 
    \textbf{a}, A PhCR with a periodic corrugation pattern, which induces strong coupling between counter-propagating optical waves resulting in frequency splitting of mode $m_0$ (angular period $\theta_0 = 2\pi/ 2m_0$).
    \textbf{b} Two pump fields aligned with resonances $P_1$ and $P_2$ provide parametric gain for the mode $S$, situated between the two pumps.
    \textbf{c}, Resonance structure and nonlinear mixing processes induced by two monochromatic pumps aligned with resonances $P_{\pm1}$, including dual-pump spontaneous four-wave mixing, single-pump spontaneous four-wave mixing (SP-SFWM), and Bragg-scattering four-wave mixing (BS-FWM).
    \textbf{d}, PhCR squeezer with the corrugation composed of two Fourier components that are aligned with modes $X_{\pm2}$, resulting in strong splitting of these modes effectively suppressing SP-FWM and BS-FWM processes.
    }
  \label{fig:concept}
\end{figure*}

However, $\chi^{(3)}$ Kerr-nonlinear systems exhibit rich four-wave mixing dynamics, with multiple nonlinear processes occurring simultaneously \cite{helt:2017, seifoory:2022}. If not suppressed, some of these processes can compromise the level of squeezing, especially for the generation of single-mode squeezed vacuum (SMSV) --- a key resource for continuous-variable quantum information processing \cite{weedbrook:2012}. Therefore, suppressing unwanted nonlinear processes remains a key challenge in such systems.
To address this, the pump lasers can be detuned from the respective microresonator modes \cite{zhao:2020, seifoory:2022}. While this mitigates parasitic effects, it reduces the achievable level of squeezing for a given pump power. Another approach involves the use of linearly \cite{zhang:2021, viola:2024} and nonlinearly \cite{menotti:2019} coupled microresonators, where careful thermal tuning of both resonators can suppress unwanted nonlinear processes and lead to efficient squeezing. 
Sharing many similarities with coupled microresonators, 
nano-corrugated photonic crystal rings (PhCR) have emerged as a new paradigm in integrated nonlinear photonics. In a PhCR forward and backward propagating waves can be coupled in a precisely controlled manner (see \reffig{fig:concept}{a}). They have enabled significant progress in soliton microcombs \cite{Yu2021, ulanov:2024, wildi:2024}, slow light \cite{lu:2022}, optical parametric oscillators \cite{black:2022}, broadband control of microresonator dispersion \cite{lucas:2023}.

Here, we demonstrate efficient single-mode quadrature squeezing based on dual-pumped spontaneous four-wave mixing process in a single Kerr-nonlinear PhCR microresonator. Critically, the coupling induced by the PhCR's nano-corrugation modifies the resonance spectrum such that unwanted nonlinear optical processes are effectively suppressed, resulting in an estimated 7.8~dB of on-chip squeezing. As opposed to a coupled-resonator approach, the nano-corrugation is static and does not need any tuning, which simplifies operation especially outside controlled laboratory environments.
Moreover, the resonator is fabricated via UV lithography in a commercial foundry process and hence compatible with scalable wafer-level fabrication, as needed for emerging quantum technologies.

\subsection*{Results}
Our squeezed light generator is based on dual-pump spontaneous four-wave mixing (DP-SFWM), where two pump lasers (P1 and P2) provide parametric gain for the signal field, located spectrally between the two pumps (mode S, see \reffig{fig:concept}{b}). Below the threshold of optical parametric oscillation, this process results in SMSV generation. However, DP-SFWM is accompanied by other nonlinear effects, including cross-phase modulation (XPM), self-phase modulation (SPM), single-pump spontaneous four-wave mixing (SP-SFWM), and Bragg-scattering four-wave mixing (BS-FWM) (see \reffig{fig:concept}{c}). While XPM and SPM induce frequency shifts that can be easily compensated for by small detunings of the pump lasers, SP-SFWM and BS-FWM introduce excess noise into the signal mode (S) if not suppressed. Therefore, optimal SMSV generation in these systems requires effective suppression of SP-SFWM and BS-FWM processes. As we explain below, we leverage a single PhCR microresonator to achieve the suppression of these processes.

The general concept of a PhCR is illustrated in \reffig{fig:concept}{a}. A spatial periodic corrugation along the resonators inner sidewall with angular corrugation period $\theta_0=pi/m_0$, creates a controllable coupling with rate $\gamma$ between the counter-propagating optical waves of 
angular (azimuthal) mode number $m_0$. The coupling of the degenerate forward-backward propagating modes creates a set hybridized modes with a frequency splitting of $2\gamma$. In our case, to suppress SP-SFWM and BS-FWM processes, we utilize a single PhCR with a periodic corrugation pattern composed of two Fourier components, designed to achieve strong splitting of modes $X_{\pm2}$ (see \reffig{fig:concept}{c,d}), so that the unwanted processes are no longer supported by the resonances. This is similar to the engineering of photonic stop bands in waveguides for photon pair generation \cite{helt:2017}.

\begin{figure*}[!ht]
  \centering
  \includegraphics[width=\textwidth]{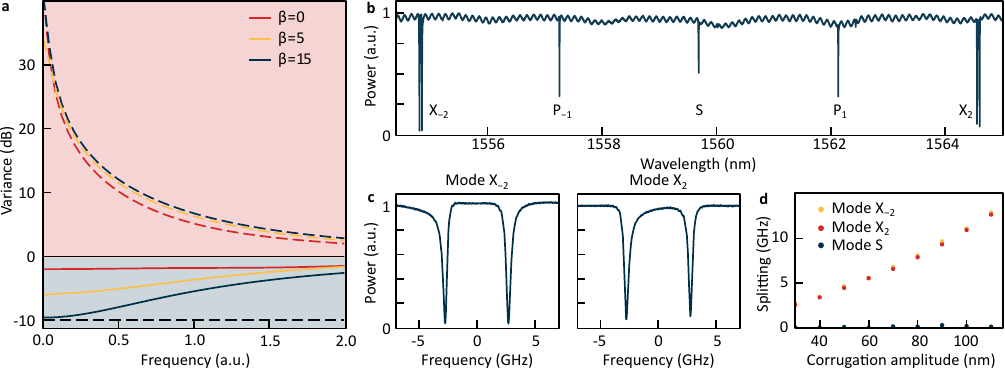}
  \caption{
    \textbf{Characterization of photonic crystal ring microresonator}. 
    \textbf{a}, Numerical simulation of squeezing and antisqueezing spectra. The red, orange, and blue lines correspond to different values of splitting of resonances $X_{\pm2}$ (see \reffig{fig:concept}{b}). The black dashed line indicates the deterministic bound, determined by the cavity outcoupling efficiency, here assumed as $\eta = 0.9$
    \textbf{b}, Transmission trace obtained during a laser wavelength scan across multiple resonances of the photonic crystal ring resonator under study.
    \textbf{c}, Zoom-in view of two split resonances $X_{\pm2}$, showing significant frequency splitting due to the corrugation pattern aligned with these modes. Other resonances remain unaffected.
    \textbf{d}, Measured frequency splitting $2\gamma$ for the two split resonances $X_{\pm2}$ and the $S$ resonance (represented by red, orange, and blue dots, respectively) as a function of the designed corrugation amplitude.
    }
  \label{fig:phcr}
\end{figure*}

To design the PhCR, as a first step, we analyse the impact of parasitic nonlinear channels on the achievable squeezing level and determine the resonance splitting required to mitigate these effects. To this end, we consider a system comprising five adjacent resonances of a high-$Q$-factor microresonator. The system is pumped by two monochromatic lasers aligned with the $P_{\pm1}$ resonances, respectively (see \reffig{fig:concept}{c}). We assume that the pumps are undepleted, the intracavity fields reach a stationary state, and all other modes remain below the oscillation threshold. Under these assumptions, the system dynamics is described by a set of coupled-mode equations governing the evolution of the creation and annihilation operators \cite{chembo:2016}. By incorporating cavity input-output relations \cite{collett:1984}, we derive the squeezing spectrum (see Methods). This framework can be generalized to accommodate an arbitrary number of modes and pump fields, allowing for the identification of maximally squeezed supermodes \cite{gouzien:2020, guidry:2023}.

The simulated squeezing spectra are shown in \reffig{fig:phcr}{a} for two pumps with equal powers and identical detunings from their respective resonances. Simulations were conducted for three different values of the normalized splitting $\beta = 2\gamma/\kappa$ of the resonances $X_{\pm2}$. Here, $\kappa$ is the cavity total loss rate, combining intrinsic cavity loss rate $\kappa_0$ and coupling rate $\kappa_\mathrm{ex}$ between the cavity and the bus waveguide. For each value of $\beta$, the pump powers and detunings are adjusted to ensure that the system remains just below the oscillation threshold. As illustrated in \reffig{fig:phcr}{a}, the antisqueezing level remains nearly constant across the simulations, while the squeezing level progressively improves with increasing $\beta$, eventually converging (at low Fourier frequencies) to the deterministic bound, which is ultimately limited by the cavity coupling efficiency $\eta = \kappa_\mathrm{ex}/\kappa$ (here we assume $\eta=0.9$). These results show that splitting the parasitic resonances (in this case, modes $X_{\pm2}$) effectively mitigates the detrimental effects of SP-SFWM and BS-FWM nonlinear processes. Without other nonlinear effects (e.g., cascaded four-wave mixing), a normalized splitting of approximately $\beta \gtrsim 15$ is sufficient to nearly achieve the optimal squeezing level, which in this case is 10~dB.

For the experiments, we fabricate a range of over-coupled resonators with varying corrugation amplitudes. At the mask design level, the corrugation amplitudes for both Fourier components are set to be equal, ensuring approximately equal splitting of the target resonances. The microresonators are fabricated using a commercial foundry process (see Methods) and have a free spectral range (FSR) of 300~GHz, corresponding to a radius of $\sim75~\mu$m. The waveguides have cross-sectional dimensions (W$\times$H) of $1.6 \times 0.8~\mu\mathrm{m}^2$, resulting in microresonators with anomalous group velocity dispersion $\beta_2 \approx - 125\,\mathrm{ps^2/km}$. The microresonator is coupled to a straight bus waveguide via a point-type evanescent coupler, and coupling to the bus-waveguide is accomplished via edge-coupling at the uncoated facets of the chip through inverse tapered spot-size converters. 

We characterize the fabricated resonators using frequency comb-calibrated laser scans \cite{delhaye:2009}, enabling the measurement of the coupling rate $\gamma$ between counter-propagating modes, the intrinsic loss rate $\kappa_0$, and the cavity coupling rate $\kappa_\mathrm{ex}$. The samples exhibit an intrinsic loss rate $\kappa_0 \approx 2\pi \cdot 55$~MHz, corresponding to an unloaded quality factor ($Q_0$) of approximately 3.5 million at a signal wavelength of $\sim 1559$~nm. The cavity coupling efficiency $\eta \approx 0.9$, close to the value assumed in the simulations presented above. We expect that through the use of pulley couplers it could potentially be increased to $\sim 0.95$.

An example transmission spectrum is presented in \reffig{fig:phcr}{b}. It shows that only the two targeted resonances (corresponding to the modes $X_{\pm2}$ in \reffig{fig:concept}{c}) exhibit significant splitting (see \reffig{fig:phcr}{c}), while other modes remain unaffected. We investigate this behavior as a function of the designed corrugation amplitude (see \reffig{fig:phcr}{d}) and confirm that both target resonances exhibit nearly identical splitting for all tested corrugation amplitudes. Based on our numerical simulations, we select a sample with frequency splitting of $\sim 5.4$~GHz (corresponds to $\beta \approx 13.5$), which is sufficient to suppress parasitic nonlinear processes effectively.

\begin{figure*}[!ht]
  \centering
  \includegraphics[width=\textwidth]{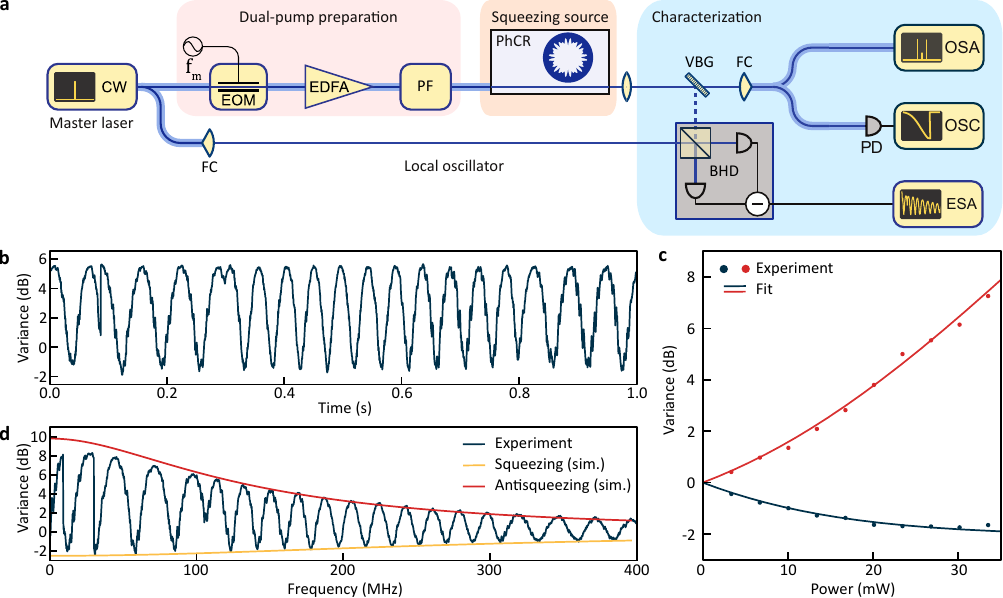}
  \caption{
    \textbf{Generation and characterization of single-mode squeezed vacuum (SMSV)}. 
    \textbf{a}, The dual-pump source is derived from a filtered, electro-optically modulated continuous-wave (CW) master laser and used to pump a photonic crystal ring resonator (PhCR), generating single-mode quadrature squeezing via dual-pump spontaneous four-wave mixing. The generated SMSV is detected by a balanced homodyne detector (BHD) and analyzed using an electronic spectrum analyzer (ESA). EOM: electro-optic modulator; EDFA: erbium-doped fiber amplifier; PF: programmable filter; VBG: volume Bragg grating filter; FC: fiber collimator; PD: photodetector; OSC: oscilloscope; OSA: optical spectrum analyzer.
    \textbf{b}, Measured quadrature variance of the SMSV at 27~mW of total pump power and 20~MHz sideband frequency.    
    \textbf{c}, Measured squeezing and antisqueezing at 20~MHz sideband frequency as a function of total on-chip pump power.
    \textbf{d}, Measured squeezing spectrum  (corrected for the BHD efficiency) at 33~mW of total pump power and simulated squeezing spectrum (see Methods). All data is taken at 2~MHz resolution bandwidth and 100~Hz video bandwidth.
    }
  \label{fig:results}
\end{figure*}

A schematic of the experimental setup for generating and measuring squeezing is shown in \reffig{fig:results}{a}. The frequency- and phase-locked dual-pump is derived from a single tunable continuous-wave (CW) master laser using electro-optic modulation, followed by amplification and filtering. The pump components are adjusted to have equal power before being coupled into the chip via a lensed fiber. The output light is collected using a large numerical aperture aspherical lens and directed to a volume Bragg grating filter, which transmits the residual dual-pump light while reflecting the generated squeezed light. The reflected light, after being overlapped with the local oscillator (LO), is directed to a free-space balanced homodyne detector (BHD). The LO is derived from the master laser to ensure coherence and frequency match with the generated squeezed light. The output signal from the BHD is then analyzed using an electronic spectrum analyzer (ESA).

To operate the setup, the electro-optic modulation frequency is adjusted to match the frequency separation of the pump modes and the CW laser is tuned so that the pumps approach their respective resonances from the blue side. Frequency tuning of the CW laser is halted when both pumps are slightly blue-detuned relative to the Kerr-shifted resonance positions, ensuring relative detuning stabilization via thermal locking \cite{carmon:2004}. This approach eliminates the need for active frequency stabilization. 

A characteristic quadrature variance trace measured in this state at 20~MHz sideband frequency as the LO phase is swept is shown in \reffig{fig:results}{b}. The directly measured squeezing (anti-squeezing) levels are $1.71\,\mathrm{dB}$ ($5.54\,\mathrm{dB}$) respectively. From these values, the overall collection and detection efficiency is determined to be $37\,\%$, consistent with the independently measured overall efficiency of approximately $40\,\%$ obtained by multiplying the efficiencies of individual components (see Methods). The remaining discrepancy is attributed to other unquantified imperfections, such as optical losses on the BHD windows, suboptimal light collection, and incomplete signal overlap with the photodiodes. The measured squeezing and anti-squeezing levels at a sideband frequency of 20~MHz, as a function of total pump power are presented in \reffig{fig:results}{c}.

Finally, we characterize the squeezing spectrum. For this purpose, we fix the total pump power at $\sim 33$~mW and sweep the offset frequency (on the ESA) and the LO phase simultaneously, to trace out the (anti-)squeezing spectra as minima and maxima of the obtained curve. The results, corrected for the BHD efficiency to exclude the detector bandwidth from consideration (see Methods), are shown in \reffig{fig:results}{d}. At low offset frequencies, we achieve $2.4$~dB squeezing and $8.3$~dB anti-squeezing. These levels correspond to $\sim 11.3$~dB of squeezing in the cavity and $\sim 7.8$~dB in the bus waveguide at $90\,\%$ outcoupling efficiency. At high offset frequencies, the levels of squeezing and antisqueezing are $0.85$~dB and $1.3$~dB, respectively. The squeezing bandwidth is limited by the cavity linewidth and exhibits good agreement with the simulated spectrum derived from the numerical model (see Methods) using the experimentally measured parameters. The slight discrepancy between the experiment and simulation is attributed to additional nonlinear processes (e.g., cascaded four-wave mixing), asymmetries in the initial pump powers, and consequently detunings \cite{delbino:2017}.

\subsection*{Conclusion}

In conclusion, we demonstrate the generation of single-mode quadrature squeezing based on DP-SFWM in a single PhCR microresonator. By engineering the mode spectrum of the PhCR, we effectively suppress parasitic nonlinear processes --- SP-SFWM and BS-FWM --- that would otherwise compromise the level of squeezing. We use numerical simulations to find the relevant design parameters and demonstrate experimentally 7.8~dB of estimated on-chip squeezing (in the bus waveguide). We anticipate, this result can be further improved to exceed 10~dB of on-chip squeezing through: (1) enhancing the cavity outcoupling efficiency via pulley couplers; (2) increasing the Kerr-nonlinear gain by introducing additional low amplitude corrugation components aligned with the pump modes $P_{\pm1}$ to achieve locally normal dispersion around the central resonance $S$; (3) improving the cavity $Q$-factor by using a wider waveguides cross-section. These improvements are all within the capability of current commercial chip foundry services.

The findings presented here highlight the potential of PhCRs for integrated quantum photonics as a scalable, chip-integrated, and CMOS-compatible solution for efficient squeezed light generation. Future work could explore integrating PhCR squeezers with on-chip optical parametric amplifiers \cite{riemensberger:2022} to mitigate downstream losses, while leveraging synthetic reflection self-injection locking \cite{ulanov:2024} may lead to ultra-compact power-efficient on-chip sources of low-operational complexity for squeezed light generation. These results open a new route to broad applications of PhCRs in quantum information processing protocols such as Gaussian boson sampling \cite{zhong:2020, madsen:2022, Arrazola2021}, coherent Ising machines \cite{inagaki:2016, mcmahon:2016, okawachi:2020}, and cluster state quantum computing \cite{larsen:2019, asavanant:2019, aghaeerad:2025}.

\subsection*{Methods}
\small
\paragraph{Numerical model.}

To simulate the squeezing spectrum, we consider a system of $N$ driven-dissipative coupled-mode differential equations describing the evolution of the annihilation ($\hat{a}_\mu$) and creation ($\hat{a}^\dagger_\mu$) operators, where $\mu$ denotes the mode number relative to the signal mode $S$ (see Fig. 1). We assume a high-$Q$-factor Kerr-nonlinear microresonator driven by multiple classical, undepleted pumps, each equally detuned from their respective cavity resonances. The corresponding intracavity fields are represented by their complex amplitudes $A_\mu$, while the other cavity modes remain below the oscillation threshold. Under these conditions, the originally quartic interaction Hamiltonian can be linearized and simplified into a quadratic form. Consequently, the evolution of the annihilation and creation operators is given by \cite{chembo:2016, guidry:2023}:
\begin{align}
\label{cme_quant}
\frac{\partial \hat{a}_\mu}{\partial t} = & \sum_{\nu} R_{\mu\nu} \hat{a}_\nu + \sum_{\nu} S_{\mu\nu} \hat{a}^\dagger_\nu  + \sqrt{\kappa_\mathrm{ex}} \hat{v}_\mathrm{ex} + \sqrt{\kappa_0} \hat{v}_0 , \\
\nonumber
R_{\mu\nu} = & - \left[ \frac{\kappa}{2} + i \left( \delta_0 + D_\mathrm{int}(\nu) \right) \right] \delta(\mu - \nu) \\
\label{Rmunu}
& + 2 i g_0 \sum_{j,k} \delta(\mu + j - \nu - k) A^*_j A_k, \\
\label{Smunu}
S_{\mu\nu} = &~i g_0 \sum_{j,k} \delta(\mu + \nu - j - k) A_j A_k.
\end{align}
Where $\delta_0$ is the pump detunings from the respective resonances (equal in our case), $g_0$ describes the cubic nonlinearity of the system, $\hat{v}_0$ and $\hat{v}_\mathrm{ex}$ are the vacuum fluctuations operators.

Equation \eqref{cme_quant} can now be rewritten in matrix form:
\begin{align}
\label{cme_matrix_temporal}
\begin{bmatrix}
    \frac{\partial \textbf{a}}{\partial t} \\
    \frac{\partial \textbf{a}^\dagger}{\partial t}
\end{bmatrix} & = 
\begin{bmatrix}
    R \quad S \\
    S^* \quad R^*
\end{bmatrix}
\begin{bmatrix}
    \textbf{a}(t) \\
    \textbf{a}^\dagger (t)
\end{bmatrix} + 
\sqrt{\kappa_0}
\begin{bmatrix}
    \textbf{v}_0 (t) \\
    \textbf{v}^\dagger_0 (t)
\end{bmatrix}
 + 
\sqrt{\kappa_\mathrm{ex}}
\begin{bmatrix}
    \textbf{v}_\mathrm{ex} (t) \\
    \textbf{v}^\dagger_\mathrm{ex} (t)
\end{bmatrix},
\end{align}
where the vectors of operators $\textbf{a}$, $\textbf{v}_0$, and $\textbf{v}_\mathrm{ex}$ are constructed from their respective annihilation operators. For example, $\textbf{a}(t) = (\hat{a}_{-n}(t), \dots, \hat{a}_{n-1}(t))^\mathrm{T}$, where $n = N/2$. Using the definitions of position $\hat{x} = (\hat{a} + \hat{a}^\dagger)/\sqrt{2}$ and momentum $\hat{p} = i (\hat{a}^\dagger - \hat{a})/\sqrt{2}$ quadratures, we can now introduce the corresponding time-dependent quadrature vectors:
\begin{align}
\textbf{Q} & = (\hat{x}_{-n} (t), \dots, \hat{x}_{n-1} (t) | \hat{p}_{-n} (t), \dots, \hat{p}_{n-1} (t))^\mathrm{T}, \\
\textbf{U} & = (\hat{x}_{0, -n} (t), \dots, \hat{x}_{0, n-1} (t) | \hat{p}_{0, -n} (t), \dots, \hat{p}_{0, n-1} (t))^\mathrm{T}, \\
\textbf{V} & = (\hat{x}_{\mathrm{ex},-n} (t), \dots, \hat{x}_{\mathrm{ex}, n-1} (t) | \hat{p}_{\mathrm{ex},-n} (t), \dots, \hat{p}_{\mathrm{ex}, n-1} (t))^\mathrm{T}.
\end{align}

In this case, Eq. \eqref{cme_matrix_temporal} is transformed as:
\begin{align}
\label{cme_quads}
\frac{\partial \textbf{Q}}{\partial t} &= \textbf{M}_q \textbf{Q} + \sqrt{\kappa_0} \textbf{U} + \sqrt{\kappa_\mathrm{ex}} \textbf{V}, \\
\textbf{M}_q &=
\begin{bmatrix}
    \Re(R + S^*)& \quad -\Im(R + S^*) \\
    \Im(R - S^*)& \quad \Re(R - S^*)
\end{bmatrix}.
\end{align}

The stationary solution of Eq. \eqref{cme_quads} can be found in the Fourier domain via matrix inversion. Combined with the cavity input-output relation, this yields:
\begin{equation}
    \textbf{Q}_\mathrm{out}(\Omega) = \sqrt{\kappa_\mathrm{ex}}(i \Omega \mathbb{I}_{2N} + \textbf{M}_q)^{-1} (\sqrt{\kappa_0} \textbf{U} + \sqrt{\kappa_\mathrm{ex}} \textbf{V}) + \textbf{V}.
    \label{Qout}
\end{equation}

Equation \eqref{Qout}, along with the commutation relations, can be used to determine the squeezing spectrum.

\paragraph{Effect of losses.}

Assuming a certain level of squeezing, defined by the squeezing parameter $r$, is achieved inside a microresonator, the variances of the squeezed and antisqueezed quadratures are given by $V_\pm = e^{\pm 2r}$. Squeezed states, and consequently the level of observed squeezing, are influenced by losses in optical channels and detector inefficiency. At the output of an optical channel with transmission $T$, the quadrature variances are expressed as:
\[
V_{\mathrm{out},\pm} = T\,V_\pm + 1 - T.
\]

Thus, by measuring both the squeezing and antisqueezing levels, it is possible to determine the original squeezing and the overall efficiency, which combines the collection and detection efficiencies.

For squeezing generated within cavities, the optimal achievable quadrature variance at the cavity output is given by:
\[
V_{\mathrm{min}} = 1 - \eta,
\]
where $\eta = \kappa_\mathrm{ex}/\kappa$ represents the cavity outcoupling efficiency \cite{scully:1997}. This efficiency ultimately limits the best possible squeezing level at the cavity output.

\paragraph{Efficiency budget.}

We carefully characterize all sources of loss that the squeezed light experiences before detection. These include the microresonator outcoupling efficiency, photonic chip outcoupling efficiency (Fresnel loss), volume Bragg grating diffraction efficiency, mode matching efficiency between the local oscillator (LO) and the signal, quantum efficiency of the balanced homodyne detector (BHD) photodiodes, optical channel loss (after chip), and the BHD's dark noise clearance (DNC). The DNC is equivalent to an optical loss and is frequency-dependent. These factors are summarized in Table~\ref{tab:eff_budget}. Taking all these contributions into account, we arrive at an overall predicted detection and collection efficiency of 0.40.

\begin{table}[!ht]
    \centering
    \begin{tabular}{lc}
        \hline
        \textbf{Source of loss} & \textbf{Value} \\
        \hline
        Ring outcoupling efficiency & 0.9 \\        
        Chip outcoupling efficiency & 0.94 \\
        Mode matching efficiency & 0.81 \\
        Optical channel efficiency (after chip) & 0.8 \\
        BHD efficiency & 0.75 \\
        BHD DNC & 0.99 - 0.78 \\
        \hline
    \end{tabular}
    \caption{Efficiency budget for squeezed light detection.}
    \label{tab:eff_budget}
\end{table}

\paragraph{Sample fabrication.} 
The samples were fabricated commercially by LIGENTEC SA using UV optical lithography.


\subsection*{Funding}
\small
This project has received funding from the European Research Council (ERC, grant agreement No 853564), from the European Innovation Council (EIC, grant agreement No 101137000) and through the Helmholtz Young Investigators Group VH-NG-1404; the work was supported through the Maxwell computational resources operated at DESY.


\printbibliography
\end{document}